\documentclass[runningheads]{llncs}
\usepackage[T1]{fontenc}
\usepackage{graphicx}
\usepackage{amsmath}
\usepackage{amssymb} 
\newcommand*{\defeq}{\stackrel{\text{def}}{=}}
\usepackage{comment} 
\begin{document}
	
\title{Specifying an Obligation Taxonomy in the 
		Non-Markovian Situation Calculus}
	
	\titlerunning{Obligation Taxonomy in the Non-Markovian Situation Calculus}
	
	\author{Kalonji Kalala \and
		Iluju Kiringa  \and
		Tet Yeap}
	
	\authorrunning{K. Kalonji et al.}
	%
	\institute{University of Ottawa, Ottawa, Ontario ON K1N 1A2, Canada \\
		\url{https://www.uottawa.ca/faculty-engineering/} \\
		\email{\{hkalo081,iluju.kiringa,tyeap\}@uottawa.ca}}
	\maketitle              
	\begin{abstract}
		Over more than three decades, the Situation Calculus has established itself as an elegant, powerful, and concise formalism for specifying dynamical domains as well as for reasoning about the effects of actions of those domains both in the world and in the mental state of the modelled agents. Moreover, it has also been     established that the preconditions of a given action and its effects may be determined entirely by the current situation alone, or they may be determined by past situations as well. When past situations are involved in determining action preconditions and effects, resulting theories are non-Markovian. Assuming a specification of actions that produce obligations, we consider using non-Markovian control in the Situation Calculus to specify different notions of obligations found in the literature. These notions have been specified using Event Calculus; but, as far as we know, they have never been specified using the Situation Calculus. The specifications in this paper yield intuitive properties that ensure the correctness of the whole endeavour.
		
		
	\end{abstract}

	\section{Introduction}\label{sec1}
	
	The Situation Calculus is a pre-eminent logical language that has been used in artificial intelligence for specifying and reasoning about dynamical systems such as robotics, database updates, control systems, simulated software agents, and computer agents. One specifies these systems by providing axioms in the language of the Situation Calculus to capture the prerequisites of actions of the domain and the effects of these actions both on the external world around the specified system as well as on the internal mental state of the dynamic  agent that is being modelled. Typically, the effects of actions are captured by fluents,  predicates whose truth values are changed as the result of performing actions. In this context, it is necessary to provide axioms that state which fluents remain unchanged during action performance. The number of such axioms is large: this is known as the frame problem in artificial intelligence. In his seminal work in \cite{DBLP:conf/birthday/Reiter91}, Raymond Reiter has proposed the so-called successor state axioms as a solution to the frame problem \cite{mcdermott1987ai} for actions that change the external world. 
	
	Reiter's solution to the frame problem relies on the Markov property, that is: the preconditions and the effects of an action are solely determined by the current situation in which the action is executed. In \cite{DBLP:conf/aaai/Gabaldon02}, Gabaldon has extended the original Reiter's solution to the Situation Calculus with the non-Markovian property; that is, the preconditions and the effects of an action may be determined by one or more of the past  situations preceding the one in which the action is executed. This extended Situation Calculus is most appropriate to specify a variety of dynamic domains that clearly showcase the non-Markovian property such as database transactions ~\cite{DBLP:conf/krdb/KiringaG03,DBLP:journals/jiis/KiringaG10}, semantics of dynamics integrity constraints ~\cite{DBLP:journals/tods/Chomicki95}, and decision-theoretic planning with reward functions that look back in the past ~\cite{BacchusBG96}. 
	
	In \cite{DBLP:conf/aaai/ScherlL93}, Scherl and Levesque extend Reiter's solution to the frame problem to actions that result in changing the state of knowledge of the reasoning agent; these actions are called {\it knowledge-producing actions}. Scherl and Levesque's solution still uses the Markovian theories to capture knowledge. 
	
	Even more sophisticated domains are emerging nowadays which require the ability to specify them  using non-Markovian theories in order to reason about them accurately. One such domain is smart contracts that have emerged in cyber physical systems as a way of enforcing formal agreements between entities that constitute these systems.  A contract is a legally enforceable agreement that contains requirements for different parties to engage in a business transaction.  Those requirements are made of {\bf obligations to be fulfilled} by the parties. In the last decade, a considerable amount of research was conducted to represent a legal contract electronically \cite{DeKruijff2017} \cite{MacLeod2005}. Such electronically codified contracts  are meant to be executed automatically  by the involved parties (contracting agents). The execution of legal  contracts must be monitored to make sure that the dealings of all parties comply with the requirements of the contract.  
	
	Any formal specification of the smart contracts mentioned above clearly calls for a explicit formal account of obligations. In~\cite{Kalonji-Kiringa-Yeap}, authors extend the Situation Calculus account of knowledge by Scherl and Levesque to the concept of obligation.      
	
	This paper uses the Situation Calculus account of obligations in the Situation Calculus presented in ~\cite{Kalonji-Kiringa-Yeap} to provide a Situation Calculus account of the taxonomy of obligations given in~\cite{DBLP:conf/ruleml/HashmiGW14}: we use non-Markovian theories of the Situation Calculus to model the various type of obligations in that taxonomy.

	The paper is organized as follows. In the next  Section~\ref{FormalPreliminaries},  we introduce the sequential temporal Situation Calculus along with deontic logic notions; here, we introduce a  running example used throughout the paper in subsection~\ref{RunningExample}.  The formalization of the obligation modality in the Situation Calculus given in ~\cite{Kalonji-Kiringa-Yeap} is summarized in Section~\ref{ObligationInSitCalc}. Section~\ref{Obligation-types} spells out the details of the formalization of the obligation types in the Situation Calculus. 
Furthermore, in Section~\ref{Properties}, we state and prove some properties of the obligation types in the Situation Calculus. We discuss the issue of the interaction between obligation violation and compensation of violated obligation in Section~\ref{Discussion-1}. Section~\ref{Discussion} discusses related work. Finally, we conclude and indicate avenues for future work in Section~\ref{Conclusion}.

	\section{Situation Calculus with Non-Markovian Control}\label{FormalPreliminaries}
	
	The Situation Calculus \cite{McCarthy63,Reiter01} is a logical framework that has been successfully employed in the formalization of a wide variety of
	dynamical systems. In Subsection~\ref{MarkovianSitCalc}, we give an overview of the Situation Calculus as formulated by Reiter \cite{Reiter01}. In Subsection~\ref{NonmarkovianSitCalc}, we complete the summary of Situation Calculus with non-Markovian control first introduced in \cite{Gabaldon02b}.
	
	\subsection{Sequential and Temporal Situation Calculus}\label{MarkovianSitCalc}
	
	The sequential and temporal Situation Calculus is a many-sorted second order logic with four sorts: actions, situations, time, and objects other than the first three sorts. Actions are first order terms that consist of an action function symbol and several arguments, the last of which is a time \cite{DBLP:conf/iclp/PintoR93}. Situations represent finite sequences of actions that are responsible for all the changes in the world or, in the case of obligations, in the mental state of the modelled agent. Fluents are properties of the world that change from situation to situation as a result of the execution of actions. 
	
	In addition to variables of the aforementioned sorts, the language includes a finite number of function symbols called action functions such as $unlock(d,t)$ to represent the actions of unlocking the door $d$ at time $t$, functions such as $do(unlock(d,t),s)$ to denote the situation resulting from the execution of the action $unlock(d,t)$ in the situation $s$, and predicates for representing fluents such as $open(d,s)$, meaning that the door $d$ is open in the situation $s$. In addition to predicate fluents, there are functional fluents that denote values that  vary from situation to situation as a consequence of executions of actions. Furthermore, the languages includes a finite number of situation independent predicates and functions. 
	The initial situation is denoted by the constant $S_0$. The function $do(a,s)$ denotes the non-empty history of actions that have been executed up to and including action $a$. Situation $S_0$ represents the starting empty history. 
	Finally, the language  also  includes special predicates $Poss$, and 
	$\sqsubset$; $Poss(a,s)$ means that the action $a$ is possible in the situation $s$, and $s \sqsubset s'$ states that the situation $s'$ is reachable from $s$ by performing some sequence of actions. 
	
	Notice that the introduction of time calls for a way of specifying times of actions as well as start times of situations that results from actions. To achieve this, first a function symbol $time(\alpha)$ to specify occurrence times of actions is introduced~\cite{DBLP:conf/kr/Reiter98}; thus, for each given action $A(\vec x, t)$, the axiomatization of the domain must include an axiom that specifies the occurrence time of that action as follows: 
	$time(A(\vec x, t)) = t$. Second, foundational axioms must include one that states that the situation $do(a,s)$ starts at the time of the action $a$, that is, $start(do(a,s))=time(a)$. 
	
	Finally, by convention in this paper, a free variable will always be implicitly bound by a prenex universal quantifier. For a complete formal description see \cite{Pirri99,Reiter01}. 
	
	
	A term $do(a_k,do(a_{k-1}, \dots, do(a_1,\sigma)\dots))$ is said to be {\it rooted at} $\sigma$ if $\sigma$ is $S_0$ or a situation variable and is the only (if any) subterm of sort situation occurring in $a_1,\dots,a_k$.	We will use the shorthand notation $do([a_1,\dots,a_k],\sigma)$ to denote the situation term $do(a_k,do(a_{k-1},\dots do(a_1,\sigma)\dots))$. 
	

	
	A Situation Calculus axiomatization of a domain, includes the
	following set of axioms:
	\begin{enumerate}
		
		\item For each action function $A(\vec{x})$, an {\sl action
			precondition axiom} of the form: 
		
		$Poss(A(x_1,\dots,x_n),s)\equiv\Pi_A(x_1,\dots,x_n,s)$
		%
		where $s$ is the only term of sort situation in $\Pi_A(x_1,\dots,x_n,s)$.
		
		\item For each fluent $F(\vec{x},s)$, a {\sl successor state axiom} of 
		the form:
		\[F(x_1,\dots,x_n,do(a,s))\equiv\Phi_F(x_1,\dots,x_n,a,s)\]
		
		where $s$ is the only term of sort situation in $\Phi_F(x_1,\dots,x_n,a,s)$.
		
		\item Unique names axioms for actions. For instance: 	
		\noindent
		$lock(d,t)\neq unlock(d,t)$.
		
		\item Axioms describing the initial situation, e.g. the initial
		database: a finite set of sentences whose only situation term is the
		constant $S_0$. 
	\end{enumerate}
	Successor state axioms are a solution to the so-called frame problem, namely the problem of providing for a small and efficient set of axioms that accounts for what is changing and what is not changing as the result of the execution of actions.  
	A set of these axioms above, together with the following set of domain independent
	foundational axioms denoted by $\Sigma$,
	\[\begin{array}{l}
		do(a_1,s_1)=do(a_2,s_2)\supset a_1=a_2\land s_1=s_2,\\
		(\forall P).P(S_0)\land (\forall a,s)[P(s)\supset P(do(a,s))]\supset (\forall s)P(s),\\
		\neg s\sqsubset S_0,\\
		s\sqsubset do(a,s')\equiv s\sqsubset s'\vee s=s',\\
	\end{array}\]
	is called a (Markovian) {\sl basic action theory}.
	Note that the foundational axioms include a characterization of the relation
	$\sqsubset$. For two situations $s_1,s_2$, $s_1\sqsubset s_2$
	intuitively means that $s_1$ is a situation that precedes $s_2$, i.e., 
	in terms of sequences of actions, the sequence $s_1$ is a prefix of $s_2$; moreover, no situation precedes $S_0$.
	
	\subsection{Non-Markovian Control in Situation Calculus}\label{NonmarkovianSitCalc}
	
	For a basic action theory without the Markov assumption, we need some preliminary definitions. These are based on those in \cite{Gabaldon02b}. 
	First, we need to introduce the following obvious abbreviations:
	\begin{align}\label{bounded quant}
		\begin{split}
			&(\exists s).\sigma'\sim\sigma''\sim\sigma \land W
			\defeq (\exists s)[\sigma'\sim\sigma''\land
			\sigma''\sim\sigma\land W]\\
			&(\forall s).\sigma'\sim\sigma''\sim\sigma \supset W
			\defeq (\forall s)[(\sigma'\sim\sigma''\land
			\sigma''\sim\sigma)\supset W]
		\end{split}
	\end{align}
	
	\noindent
	where $\sim$ stands for $\sqsubset$, $=$ or $\sqsubseteq$, 
	and variable $s$	appears in $\sigma''$. If $\sigma'$ is
	$S_0$ then we may write $(\exists s:\sigma''\sim\sigma)W$ and
	$(\forall s:\sigma''\sim\sigma)W$ instead.

	\begin{definition}[Bounded Formulas]\label{bounded formulas}
		For $n\geq 0$, let $\sigma$ be a term
		$do([\alpha_1,\dots,\alpha_n],\lambda)$ rooted at $\lambda$. The
		formulas of the Situation Calculus {\sl bounded} by $\sigma$ are the
		smallest set of formulas such that:
		\begin{enumerate}
			\item [1] If $W$ is an atom whose situation terms are all
			rooted at $\lambda$, then $W$ is bounded by $\sigma$.
			\item [2] If $W',W''$ are formulas bounded by situation 
			terms rooted at $s$ and $\lambda$, respectively, then 
			$(\exists s).\sigma'\sim\sigma''\sim\sigma \land W$ and
			$(\forall s).\sigma'\sim\sigma''\sim\sigma \supset W$
			are formulas bounded by $\sigma$, where $\sigma''$ 
			is rooted at $s$ and $W=(\neg)(W'\land W'')$.
			\item [3] If $W_1,W_2$ are formulas bounded by situation 
			terms rooted at $\lambda$, then $\neg W_1$, $W_1\land W_2$ 
			and $(\exists v)W_1$, where $v$ is of sort {\sl action} 
			or {\sl object}, are formulas bounded by $\sigma$.
		\end{enumerate}
	\end{definition}
	
	The set of formulas {\sl strictly bounded} by $\sigma$ is similarly
	defined by requiring in item~(1) above that all situation terms of $W$ be
	subterms of $\sigma$, in item~(2) that $W'$ be strictly
	bounded by a subterm of $\sigma''$ and $W''$ by a subterm of $\sigma$;
	and in item~(3) that $W_1,W_2$ be strictly bounded by subterms of $\sigma$.

	\medskip
	Non-Markovian basic action theories differ from Markovian ones by using action preconditions and successor state axioms in which preconditions and effects of actions may depend on past situations, and not only on the current
	one. 
	
	Formally, this means that the formula $\Pi_A(x_1,\dots,x_n,s)$ which is the right hand side of an action precondition axiom  will be a formula  
	bounded by the situation term $s$ in which $Poss$ is not mentioned 
	and that may refer to past situations of the kind given in  abbreviations~(\ref{bounded quant}).  Moreover, this means also that the formula         
	$\Phi_F(x_1,\dots,x_n,a,s)$, which is the right hand side of a successor state axioms, will be a formula strictly bounded by $s$.
	
	
	For basic action theories with the Markov assumption, Pirri \& Reiter
	\cite{Pirri99} define a {\sl regression} mechanism that
	takes a non-Markovian Situation Calculus sentence and, under certain restrictions on the form of this sentence, compiles it into an equivalent sentence that mentions no other situation term than $S_0$. This allows proving sentences without appealing to the foundational axioms $\Sigma$. This regression operator was generalized for non-Markovian theories in \cite{Gabaldon02b}. For the sequel of this paper which does not include reasoning, we do not need caring about the regression mechanism.

\subsection{Running Example: Opening a Door} \label{RunningExample}


Consider a door opening domain where the agent is opening a door. Agents  open and close the door depending on whether certain conditions are met. To start, we enumerate some of the actions and fluents and situation independent predicate and functions of the domain. {\bf Primitive actions} are: 
$unlock(d,t)$,  
$pressButton(d,c,t)$, 
$lock(d,t)$, 
$moveTo(d)$, and  
$notify(m,t)$. 
As for {\bf Fluents}, we have: 
$open(d,s)$,  
$locked(d,s)$,  
$notifiedManager(s)$, and 
$at(d,s)$. Finally, {\bf situation independent predicates and function} are:  
$manager(m)$, and  
$door(d)$. 
All these actions, fluents, functions, and predicates are intuitively understandable, except the following: 
\\[.7ex]
{\bf Primitive Actions}.
\\[-2ex]
\begin{itemize}    
	\item $pressButton(d,E,t)$: press the button to open the door $d$ with
	credential $E$ at time $t$; $E$ is a constant meaning "Employee".  
	\item $notify(m,t)$: notify the manager $m$ of the locking of the door at 
	time $t$.    	 	
\end{itemize}  
{\bf Fluents}.
\begin{itemize} 
	\item $notifiedManager(s)$: functional fluent meaning the manager has been 
	notified of the locking of the door.   
	\item $at(d,s)$: the agent is at the door $d$ in situation $s$.
\end{itemize} 
{\bf Action Precondition Axioms}. There is one for each action function $A({\vec x,t})$ listed above: 
\begin{align}  
	\begin{split}\label{APA-unlock}  
		Poss(&unlock(d,t),s) \equiv  door(d)\land at(d,s) \land 
		locked(d,s), 
	\end{split}\\
	\begin{split}\label{APA-lockDoor}
		Poss(&lock(d,t),s)  \equiv 
		open(d,s) \land at(d,s) \land door(d), 
	\end{split}\\ 
	\begin{split}\label{APA-moveTo}
		Poss(&moveTo(d,t),s)\equiv  true, 
	\end{split} \\
	\begin{split}\label{APA-pressButto}  
		Poss(&pressButton(d,c,t),s) \equiv  
		at(d,s)\land door(d)\land c = E. 
	\end{split}\\
	\begin{split}\label{APA-NOTIFY}
		Poss(&notify(m,t),s)\equiv  true.
	\end{split}
\end{align}
{\bf Successor State Axioms}. There is one such axiom for each $(n+1)$-ary relational fluent $F$, and there is one for each $(n+1)$-ary functional fluent $f$: 
\begin{align} 
	\begin{split}\label{SSA-openDoor}
		&open(d, do(a,s)) \equiv 
		(\exists t,c). a = pressButton(d,c,t) \lor  \\
		&\hspace{2.5cm}open(d,s) \land a \not = lock(d,t),
	\end{split}\\  
	\begin{split}\label{SSA-lockedDoor}
		&locked(d,do(a,s)) \equiv 
		(\exists t)a = lock(d,t) \lor (locked(d,s) \land \\
		&\hspace{2.5cm}\neg (\exists t', c)(c = E \land 
		a = pressButton(d,c,t'))),
	\end{split} \\
	\begin{split}\label{SSA-at}
		&at(d,do(a,s)) \equiv \\
		&\hspace{1cm} (\exists t) a=moveTo(d,t) \lor
		at(d,s) \land \neg (\exists d',t') moveTo(d',t'),
	\end{split} \\
	\begin{split}\label{SSA-notifiedManager}
		&notifiedManager(do(a,s))=m \equiv \\
		&\hspace{1.5cm} (\exists t) (manager(m) 
		\land a = notify(m,t)) \lor 
		notifiedManager(s)=m.
	\end{split}
\end{align}
Assume that we have successor state axioms for the other fluents listed earlier as well.


\section{Obligations in the Situation Calculus}\label{ObligationInSitCalc}

\subsection{A Deontic Fluent for Expressing Obligations}\label{deonticFluent}
Before expressing any taxonomy of obligation types such the one provided in \cite{DBLP:conf/ruleml/HashmiGW14} in the Situation Calculus, one must have already expressed the obligation modality in the Situation Calculus. That is, we must assume a solution for an embedding of the obligation modality in the Situation Calculus. In the sequel, we want to summarize an embedding proposed in~\cite{Kalonji-Kiringa-Yeap}.

The idea of using the Situation Calculus to specify obligations has been floated by a previous source:  in \cite{DBLP:conf/deon/DemolombeH04}, Demolombe and Herzig give the first formulation of Deontic Logic in the Situation Calculus to capture an agent's set of obligations ~\cite{DBLP:journals/aepia/DemolombeP09}. They use Scherl and Levesque's approach \cite{DBLP:conf/aaai/ScherlL93} of not directly resorting to some modal logic and by introducing modal logic possible worlds as the situations of the Situation Calculus. This allows one to represent obligations directly in the language of the Situation Calculus by using an appropriate first-order representation of the accessibility relation of semantics of modal logics: this accessibility relation is captured as a fluent $O(s',s)$ to be read as: “$s’$ is deontically accessible from $s$”.    

In~\cite{Kalonji-Kiringa-Yeap}, authors go a step further than in \cite{DBLP:conf/deon/DemolombeH04} by combining Deontic modalities with time in the Situation Calculus to  give an account of actions that produce obligations and extend Reiter's regression mechanism to the new obligation setting. Thus, the authors of ~\cite{Kalonji-Kiringa-Yeap} account for the extension of Reiter's solution to the frame problem for "world" actions of an application domain to actions that change the state of an agent's obligations. 

Like in~\cite{DBLP:conf/deon/DemolombeH04}, the following proposal is used in~\cite{Kalonji-Kiringa-Yeap} to represent that, in situation $s$,  an agent is obliged to bring about $\phi$:  
\begin{align}\label{OBLG-DEFINITION-1}
	(\forall s'). \textit{\textit{O}} (s',s) \supset \phi\left[ s'\right],   
\end{align}
where $\phi[s']$ represents the formula $\phi$ with  situation arguments added recursively to fluents that occur in $\phi$. So, to express that it is obligatory to have the $door$ locked by an agent, we will write: 
\begin{equation} \label{OBLG-DEFINITION-1} 
	\begin{aligned}                        
		(\forall s'). O(s',s) \supset locked(d, s').
	\end{aligned}
\end{equation} 
In representing the obligation modality, the notation $\textbf{Oblg}(\phi, s)$ is used to capture the fact that, in situation $s$, it is obligatory to the modelled agent that $\phi$ is true:  
\begin{align}\label{OBLG-DEFINITION-2}
	\textbf{Oblg}(\phi, s)  \defeq (\forall s'). \textit{\textit{O}}
	(s',s) \supset \phi\left[ s'\right].   
\end{align}
The formula $\phi$ used in the abbreviation $\textbf{Oblg}(\phi, s)$  represents a  formula obtained from a Situation Calculus formula by recursively suppressing its situation arguments. Conversely, $\phi[s]$ represents a Situation Calculus formula obtained by recursively restoring its suppressed situation arguments.  

This definition (\ref{OBLG-DEFINITION-2}) as well as the formula (\ref{OBLG-DEFINITION-1}) are based on the semantic condition $(C.O^+)$  given in \cite{Hintikka70} as Kripke semantics for the obligation modality.  
For example, $\textbf{Oblg}\big(locked(d), s \big)$  expands as follows: 
\begin{equation}  
	\begin{aligned}                     
		\textbf{Oblg}(locked(&d),s)\defeq 
		(\forall s'). O(s',s) \supset locked(d,s').
	\end{aligned}
\end{equation}

\subsection{Obligation-Producing Actions}\label{ObligationProducingActions}

With a formal concept of obligation in the Situation Calculus in hand, we can now link obligations with actions performed by agents. Among these, some do affect what is happening in the world and some others do not, but rather do affect an agent's state of obligations. We call the later obligation-producing actions, by reference to knowledge-producing actions introduced by Scherl and Levesque in \cite{DBLP:conf/aaai/ScherlL93}. At the atomic level, obligation-producing actions are of two kinds: those actions whose effect is to make some (atomic) formula obligatory, and those whose effects is to make the denotation of some term obligatory. 

As an example of the first kind of obligation-producing actions, the ground action  
$unlock(D,10)$  
executed by the agent in situation $S_0$ makes the ground atomic formula $locked(D,do(unlock(D,10),S_0))$ obligatory. 
In other words, the sentence $\textbf{Oblg}(locked(D),do(unlock(D,10),S_0))$ is made true by the execution of the action $unlock(D,10)$.
In our running example, $unlock(d,t)$ is an obligation-producing action that creates the obligation for the agent to subsequently get the door locked. That is, by executing the action $unlock(D,10)$ in situation $S_0$, the agent has the obligation to make sure that in some situation $S$ following the situation $do(unlock(D,10),S_0)$, the formula $locked(D,S)$ is true by virtue of an action executed by the agent to make $locked(D,S)$ true.


In general, we assume that there is a provision of (finitely many) obligation-producing actions $a_{F_i}(\vec x_i,s)$ where $i= 1 \ldots m$, and that for each one of them, there is a  fluent $F_i(\vec x_i,s)$, $i= 1 \ldots m$, of the domain that is made obligatory in situation $do(a_{F_i}(\vec x_i),s)$ upon the execution of $a_{F_i}(\vec x_i,s)$ in situation $s$. 


For the second kind of obligation-producing actions, we assume that 
there is a provision of (finitely many) such actions $a_{f_j}(\vec x_j)$ where $j= 1 \ldots n$, and that for each one of them, there is a functional fluent $f_j(\vec x_j)$, $j= 1 \ldots n$, of the domain whose denotation is made obligatory to the agent.

\subsection{The Frame Problem for Obligation-Producing Actions}\label{FrameProblemForObligations} 

Suppose the agent executes the action $set(E,20)$ in $S_0$ where no obligation holds and the door is $locked(D,S_0)$ is true. Then,  $credential(E,do(set(E,20),S_0))$ holds, and no new obligation is introduced. Now, if $unlock(D,30)$ is executed, the sentence $open(D,do(unlock(D,30),do(set(E,20),S_0)))$ holds and the agent has the obligation of subsequently locking the door. Finally, the execution of the action $lock(D,40)$ will stop the obligation for the agent to get the door locked.

What the above consideration shows is that we have three sorts of actions. The first sort is made of a provision of ordinary actions that do not produce any obligation with respect to the agent and do not defeat any existing obligations. The second sort is made of ordinary actions that do not produce any obligation, but they stop existing obligation. Finally, the third sort is made of obligation-producing actions.

To start, suppose that a deontic agent is located in a situation $s$. We can imagine several other infinitely many situations $s'_1, s'_2, s'_3, \ldots$, which are deontic alternatives to $s$. Furthermore, suppose that the deontic agent performs some action $a$ in $s$ and therefore lands in the successor situation $do(a,s)$. We now wonder what are the deontic alternatives to $do(a,s)$, and how these alternatives are related to the situations $s'_1, s'_2, s'_3, \ldots$. We must come up with successor state axioms for the three sorts of actions identified above by spelling out how an action $a$ affects the fluent $O$. Proposals for such successor state axioms are given in~\cite{DBLP:conf/deon/DemolombeH04}, and in ~\cite{Kalonji-Kiringa-Yeap}: we refer the reader to these works for details.

\section{Obligation Types in the Situation Calculus}\label{Obligation-types}

In ~\cite{governatori2013abstract}~\cite{DBLP:conf/assri/HashmiGW13}, the author has proposed a taxonomy of obligation types which the authors of ~\cite{DBLP:conf/ruleml/HashmiGW14} have revisited and modelled in the Event Calculus~\cite{DBLP:conf/slp/KowalskiS94}\cite{kowalski1989logic}\cite{Shanahan99}\cite{Mueller2015}. The modelled obligation types are: punctual, persistent (with maintenance and achievement variants), perdurant, and compensated.


In each of the subsections in the sequel of this section, we first introduce a given obligation type by providing its definition taken from the work in   ~\cite{DBLP:conf/ruleml/HashmiGW14}. Then we provide abbreviations that model these various types in the non-Markovian Situation Calculus.  


To start, we introduce two functions that are used in the definitions of the obligation types. Suppose that $\cal{L}$ is the language of the Situation Calculus. A {\bf State} is defined as a function $\mathbb{N}\mapsto 2^{\cal{L}}$, such that, for a given time point $t$, $State(t)$ is the set of all formulas that are valid at time $t$. That is, $\textbf{State}(t) = \{\phi(s): \phi(s) \text{~and~} time(s)=t\}$. 
Furthermore, an obligation is said to be in force at time $t$ if 
$\textbf{Oblg}(\phi, s)$ and $start(s) = t$. Thus, {\bf Force} is defined as a function $\mathbb{N}\mapsto 2^{\cal{L}}$ such that, for a given time point $t$, 
$Force(t)$ is the set of all obligations that are valid at time $t$. That is, 
$\textbf{Force}(t) = \{\phi(s): \textbf{Oblg}(\phi, s) \text{~and~} time(s)=t\}$.

%

\subsection{Punctual Obligation}\label{PunctualObligation}

An obligation $\phi$ is punctual at a time point $t \in \mathbb{N}$, if and only if we have: for  time point  $t-1$, $\phi \not\in \textbf{Force}(t-1)$; for  time point $t+1$, $\phi \not\in \textbf{Force}(t+1)$; and $\phi \in \textbf{Force}(t)$. The abbreviation below expresses the punctual obligation in the Situation Calculus in terms of situations.  
\begin{align}\label{punctual-obligation-1}
	\begin{split} 
		\textbf{OblgAt}(\phi,s) \defeq 
		&(\forall s',a_1) \left[s=do(a_1,s')  
		\supset \neg \textbf{Oblg}(\phi,s')\right] \wedge \\
		&(\forall s'',a_2) \left[s'' = do(a_2,s) 
		\supset \neg \textbf{Oblg}(\phi,s'') \right] \wedge \textbf{Oblg}(\phi,s).
	\end{split}
\end{align}
Intuitively, $\textbf{OblgAt}(\phi,s)$ in the abbreviation ~(\ref{punctual-obligation-1}) means that the formula $\phi$ is obligatory at the situation $s$ if and only if $\phi$ is obligatory at the situation $s$; moreover, $\phi$ is neither obligatory at the situation immediately before the situation $s$, nor at the situation immediately after $s$. 

A punctual obligation that is not fulfilled exactly at situation  $s$ is said to be violated. A violation of a punctual obligation $\phi$ in situation $s$ means that at $s$, \(\phi\) is not true; formally, this means that we  have: $\neg\phi[s]$. To specify the fact that a given obligation $\phi$ is violated in situation $s$, we need a predicate $violatedType(\phi, s)$, where the substring $Type$ encodes the type of the the obligation being violated; $violatedType(\phi, s)$ abbreviates an appropriate formula that formally spells out the conditions under which the violation occurs. The following abbreviation spells out the violation condition for the punctual obligations:
\begin{align}\label{punctual-obligation-violation-1}
	\begin{split} 
		violatedAt(\phi,s) \defeq \neg\phi[s].
	\end{split}
\end{align}

The specification~\ref{punctual-obligation-1} above is given in terms of situations. We can easily translate it into a specification in terms of time points as follows:  
\begin{align}\label{punctual-obligation-2}
	\begin{split} 	
		&\textbf{OblgAt}(\phi,t) \defeq \\
		&\hspace{.5cm}(\exists s)[\textbf{Oblg}(\phi,s)
		                                  \land start(s) = t \land \\   
	    &\hspace{1.5cm}(\forall s',a_1) \left[s=do(a_1,s')    
		\supset \neg \textbf{Oblg}(\phi,s')\right] \wedge \\
		&\hspace{1.5cm}(\forall s'',a_2) \left[s'' = do(a_2,s) 
		\supset \neg \textbf{Oblg}(\phi,s'') \right] ].
	\end{split}
\end{align}
Intuitively, $\textbf{OblgAt}(\phi,t)$ in the abbreviation ~(\ref{punctual-obligation-2})    means that the formula $\phi$ is obligatory at time $t$ if and only if $\phi$ is obligatory at some situation $s$ whose start time is $t$; furthermore, $\phi$ is neither obligatory at the situation immediately before the situation $s$, nor at the situation immediately  after $s$.

\subsection{Persistent Obligation}

An obligation $\phi$ is persistent between time points $t_1 \in \mathbb{N}$ and $t_2 \in \mathbb{N}$, if and only if we have: $t_1 < t_2$; for $t' =  t_1-1$, $\phi \not\in \textbf{Force}(t')$; for  $t'' = t_2+1$, $\phi \not\in \textbf{Force}(t'')$;  and for all time points $t'''$ such that $t_1 \leq t''' \leq t_2$, $\phi \in \textbf{Force}(t''')$. A persistent obligation that is not fulfilled at least one of the points in the interval $[t_1, \ldots, t_2]$ is said to be violated.  The Situation Calculus specification of the persistent obligation in terms of situations is expressed in the following  abbreviation:  
\begin{align}\label{OBLIGATION-Persistent-1}
\begin{split} 
	\textbf{OblgPersist}(\phi,s_1, s_2) & \defeq \\
	&(\forall s',a_1)  \left[s_1=do(a_1,s') 
	\supset \neg \textbf{Oblg}(\phi,s') \right] \wedge \\
	&(\forall s'', a_2) \left[s'' = do(a_2,s_2)  
	\supset \neg \textbf{Oblg}(\phi,s'') \right] \wedge \\
	&(\forall s''') \left[(s_1 \sqsubseteq  s''' \sqsubseteq  s_2 )                        
	\supset \textbf{Oblg}(\phi,s''') \right]
\end{split}
\end{align}	
In the abbreviation ~(\ref{OBLIGATION-Persistent-1}), 
$\textbf{OblgPersist}(\phi,s_1, s_2)$ means that the formula $\phi$ is obligatory between situations $s_1$ and $s_2$ if and only if $\phi$ is obligatory between situations $s_1$ and $s_2$ inclusively; moreover, $\phi$ is neither obligatory at the situation immediately before the situation $s_1$, nor at the situation immediately after $s_2$.


In terms of time points, we have the following specification:  
\begin{align}\label{OBLIGATION-Persistent-2}
\begin{split} 
	\textbf{OblgPersist}(&\phi,t_1, t_2) \defeq \\
	(\exists s_1, s_2)[
	&(\forall s') [s_1=do(a_1,s') \wedge start(s_1) = t_1 
	\supset \neg \textbf{Oblg}(\phi,s')] \wedge \\
	&(\forall s'')[s'' = do(a_2,s_2)  \wedge start(s_2) = t_2
	\supset \neg \textbf{Oblg}(\phi,s'')] \wedge \\
	&(\forall s''')[((s_1 \sqsubseteq  s''')  
	\wedge (s''' \sqsubseteq  s_2 ) \wedge 	\\	
	&\hspace{1.5cm} start(s_1) = t_1  \wedge start(s_2) = t_2 ) 
	\supset \textbf{Oblg}(\phi,s''')]] 
\end{split}
\end{align}	
Intuitively, $\textbf{OblgPersist}(\phi,t_1, t_2)$ in the abbreviation ~(\ref{OBLIGATION-Persistent-2}) is to be read in a manner similar to the reading of the abbreviation ~(\ref{punctual-obligation-2}), mutatis mutandis.   

\subsubsection{Achievement Obligation}

Persistent obligations have two variants, namely achievement and maintenance obligations. This section specifies the achievement obligations and the next one deals with maintenance obligations.  

A formula $\phi$ is an achievement obligation between time points $t_1 \in \mathbb{N}$ and $t_2 \in \mathbb{N}$, if and only if we have: $t_1 < t_2$;    and for all time points $t'$ such that $t_1 \leq t' \leq t_2$, $\phi \in \textbf{Force}(t')$.  The following abbreviation captures achievement obligations in the Situation Calculus in terms of situations: 
\begin{align}\label{OBLIGATION-Achievement-1}
\begin{split} 
	\textbf{OblgAchieve}( \phi,s_1, s_2) \defeq  
	(\forall s')\left[(s_1 \sqsubseteq  s' \sqsubseteq  s_2 )                        
	\supset \textbf{Oblg}(\phi,s') \right]
\end{split}
\end{align}	

There are two sub-variants of achievement obligations. The first sub-variant is called pre-emptive obligation; intuitively, the latter is an achievement  obligation that can be fulfilled before it becomes in force in the interval $[t_1, t_2]$; that is, we have at least a $t_3$ such that $t_3 \leq t_2$,  and $\phi \in \textbf{State}(t_3)$. The abbreviation~(\ref{OBLIGATION-Achievement-Preemptive-1}) captures pre-emptive achievement obligations in the Situation Calculus in terms of situations.  
\begin{align}\label{OBLIGATION-Achievement-Preemptive-1}
\begin{split} 
\textbf{OblgPreemptive}( \phi,s_1, s_2) & \defeq \\
& (\forall s')\left[(s_1 \sqsubseteq  s' \sqsubseteq  s_2 )                        
	\supset \textbf{Oblg}(\phi,s') \right]\land \\
& (\forall s'')(\exists s_3)\left[s''\sqsubseteq s_3 \sqsubseteq s_2 
	\land \phi[s_3]\right]
\end{split}
\end{align}	
The second sub-variant of achievement obligations is called non pre-emptive obligation. Intuitively, the latter is an achievement  obligation that needs not be fulfilled before it is in force in the interval $[t_1, t_2]$. 
The specification of non pre-emptive obligations in terms of situations is as follows: 
\begin{align}\label{OBLIGATION-Achievement-NonPreemptive-1}
\begin{split} 
	\textbf{OblgNonPreemptive}(\phi,s_1, s_2) & \defeq \\
	& (\forall s')\left[(s_1 \sqsubseteq  s' \sqsubseteq  s_2 )                        
	\supset \textbf{Oblg}(\phi,s') \right] \land \\
	& (\exists s_3)\left[s_1\sqsubseteq s_3 \sqsubseteq s_2 
	\land \phi[s_3]\right]  
\end{split}
\end{align}

We now turn to abbreviations that spell out violation conditions for the various types of obligations. Consider a preemptive obligation  $\phi$ to be fulfilled in the interval $[s_1, \ldots, s_2]$. The following abbreviation spells out the violation condition for preemptive obligations:
\begin{align}\label{Preemptive-obligation-violation-1}
\begin{split} 
 \textbf{violatedPreemptive}(\phi,s_1,s_2) \defeq 
  (\forall s) [s \sqsubseteq s_1 \sqsubseteq s_2 \land \neg\phi[s]].
\end{split}
\end{align}
Informally, the abbreviation~(\ref{Preemptive-obligation-violation-1}) stipulates that a preemptive obligation  $\phi$ is said to be violated iff it is not fulfilled at least one of the situations in the interval $[s_1, \ldots, s_2]$.  
 
Consider a non-preemptive obligation  $\phi$ to be fulfilled in the interval $[s_1, \ldots, s_2]$. The violation condition for non-preemptive obligations is given as follows:
\begin{align}\label{NonPreemptive-obligation-violation-1}
\begin{split} 
 \textbf{violatedNonPreemptive}(\phi,s_1,s_2) \defeq 
  (\forall s) [s_1 \sqsubseteq s \sqsubseteq s_2 \land \neg\phi[s]].
\end{split}
\end{align}

\subsubsection{Maintenance Obligation} 

We now turn our attention to the maintenance obligation. A formula $\phi$ is a maintenance obligation between time points $t_1 \in \mathbb{N}$ and $t_2 \in \mathbb{N}$, if and only if we have: $t_1 < t_2$;    and for all time points $t'$ such that $t_1 \leq t' \leq t_2$, $\phi \in \textbf{Force}(t')$ and 
$\phi \in \textbf{State}(t')$.  The following abbreviation captures maintenance obligations in the Situation Calculus in terms of situations: 
\begin{align}\label{OBLIGATION-Maintenance-1}
\begin{split} 
	\textbf{OblgMaintenance}( \phi,s_1,& s_2) \defeq  \\
	(\forall s')[(s_1 & \sqsubseteq  s' \sqsubseteq  s_2 )                        
	\supset [\textbf{Oblg}(\phi,s') \land \phi[s']]]
\end{split}
\end{align}	
The following abbreviation specifies the violation condition for a  maintenance obligation $\phi$ in the interval $[s_1, \ldots, s_2]$:
\begin{align}\label{Maintenance-obligation-violation-1}
\begin{split} 
 \textbf{violatedMaintenance}(\phi,s_1,s_2) \defeq 
  (\exists s) [s \sqsubseteq s_1 \sqsubseteq s_2 \land \neg\phi[s]].
\end{split}
\end{align}

\subsection{Perdurant Obligation} 

We finally deal with the perdurant obligation. A formula $\phi$ is a perdurant obligation with deadline $d$ between time points $t_1 \in \mathbb{N}$ and $t_2 \in \mathbb{N}$, if and only if we have:      $t_1 < d < t_2$;    and for all $t'$ such that $t_1 \leq t' \leq t_2$, $\phi \in \textbf{Force}(t')$.  The following abbreviation captures perdurant obligations in the Situation Calculus in terms of situations: 
\begin{align}\label{OBLIGATION-Perdurance-1}
\begin{split} 
	\textbf{OblgPerdurant}( \phi,s_1, d, s_2) \defeq  
   (\forall s)[(s_1  \sqsubseteq d 
                                  \sqsubseteq  s \sqsubseteq  s_2 )                        
	\supset \textbf{Oblg}(\phi,s)]
\end{split}
\end{align}	
The violation condition for a  perdurant obligation $\phi$ in the interval $[s_1, \ldots, s_2]$ with deadline $d$ is expressed in the following:
\begin{align}\label{Perdurance-obligation-violation-1}
\begin{split} 
 \textbf{violatedPerdurance}(\phi,s_1,d,s_2) \defeq 
  (\forall s) [s_1\sqsubseteq s \sqsubseteq d \sqsubseteq s_2 
     \supset \neg\phi[s]].
\end{split}
\end{align}

\section{Correctness}\label{Properties}

This section turns to the correctness of the formalization of the obligation types that were modelled in the Situation Calculus in Section~\ref{Obligation-types}. For each one of the obligation types, we modelled the obligation itself as well as its violation.  Therefore, for each obligation type, we need to show that the right-hand sides of the abbreviations that model the obligations and their violations indeed do reflect the informal definitions of the obligations and their violations. 
In the proofs, we need to relate the main predicates used in the formalization to the functions $Force$ and $State$ as follows:   
\begin{itemize}
\item (R1): \( \textbf{Oblg}(\phi,t)\) holds \(\iff \phi \in \textbf{Force}(t)\) 
\item (R2): \(\neg \textbf{Oblg}(\phi,t)\) holds \(\iff \phi \not \in 
                                                        \textbf{Force}(t)\)
\item (R3): $\phi[t]$ holds \(\iff \phi \in \textbf{State}(t)\) 
\end{itemize}
Note that these correlations can be expressed in terms of situations as well, e.g., 
\begin{itemize}
\item (R1): \( \textbf{Oblg}(\phi,s)\) holds 
                                       \(\iff \phi \in \textbf{Force}(s)\).
\end{itemize}
\begin{lemma}(\textbf{Punctual Obligation})
If \( \textbf{OblgAt}(\phi,t)\) is true, then \(\phi\) is a punctual obligation in force at time $t$, not in force neither at time $t-1$ nor at time $t+1$; that is, \(\phi\in \textbf{Force}(t)\), \(\phi\not\in Force(t-1)\), and \(\phi\not\in \textbf{Force}(t+1)\).
\end{lemma}
\begin{proof}
Assume the semantics of punctual obligations, which is provided by the combination of the following conditions : \textbf{(1)}: \(\phi \in \textbf{Force}(n)\);  \textbf{(2)}: \(\phi \not \in \textbf{Force}(n - 1)\); and \textbf{(3)}: \(\phi \not \in \textbf{Force}(n + 1)\). Now suppose that, for a fixed formula $\Phi$ and a fixed time point $T$, \( \textbf{OblgAt}(\Phi,T)\) holds. From the right hand side of the abbreviation (\ref{punctual-obligation-2}), with appropriate substitutions and skolemization, we get the following conjunction:
\begin{align}\label{punctual-obligation-RHS-conjuncts-2}
\begin{split} 	 
             [&\textbf{Oblg}(\Phi,S)\land start(S) = T]\land \\   
              & (\forall s',a_1) \left[S=do(a_1,s')    
		         \supset \neg \textbf{Oblg}(\Phi,s')\right] \wedge \\
              & (\forall s'',a_2) \left[s'' = do(a_2,S) 
		         \supset \neg \textbf{Oblg}(\Phi,s'') \right] ].
\end{split}
\end{align} 
We must show that the three conditions \textbf{(1)}, \textbf{(2)}, and \textbf{(3)} hold. \\[.2cm]
Condition \textbf{(1)}: From the first conjunct 
in the formula~\ref{punctual-obligation-RHS-conjuncts-2} we get that 
$\textbf{Oblg}(\Phi,S)$ and $start(S) = T$ hold; therefore, by the correlation R(1), we have $\Phi(S)\in \textbf{Force}(S)$.\\[.2cm]
Condition \textbf{(2)}: From the second conjunct 
in the formula~(\ref{punctual-obligation-RHS-conjuncts-2}) and by fixing $a_1$ and $s_1$, we get that 
\(\left[S=do(A_1,S')    
		         \supset \neg \textbf{Oblg}(\Phi,S')\right].\) 
That is, suppose we have $S=do(A_1,S')$, that is, we have the immediate predecessor situation $S'$ of S; then we conclude that 
$\neg \textbf{Oblg}(\Phi,S')$. Hence, by (R2), we have that $\Phi \not\in \textbf{Force}(S')$.  \\[.2cm]
Condition \textbf{(3)}: By a similar argument as for Condition \textbf{(2)}, The third conjunct in the formula~\ref{punctual-obligation-RHS-conjuncts-2}) 
yields 
\(\left[S'' = do(A_2,S) 
		         \supset \neg \textbf{Oblg}(\Phi,S'') \right],\) 
and, from there, $\Phi \not\in \textbf{Force}(S'')$. 
$\Box$
\end{proof}

\begin{lemma}(\textbf{Violation of Punctual Obligation})
If  $\textbf{violatedAt}(\phi,s)$ is true, then \(\phi\) is a punctual obligation in force at situation $s$, that is, \(\phi\in \textbf{Force}(s)\), and $\phi \not\in \textbf{State}(s)$.
\end{lemma}
\begin{proof}
For the violation of the punctual obligation, the right hand side of the abbreviation~(\ref{punctual-obligation-violation-1}) implies that \(\phi\) is not $\textit{true}$; that is, \( \neg \phi[s]\). Hence, by the  correlation (R3), we have \(\phi \not \in \textbf{State}(s)\).
$\Box$
\end{proof}
 
\begin{lemma}(\textbf{Perdurant Obligation})
If \( \textbf{OblgPerdurant}(\phi,s_1,d,s_2)\) holds, then \(\phi\) is a perdurant obligation in force in all situations $s$ in the interval $[s_1 \dots s_2]$, that is, for all $s \in [s_1 \dots s_2]$, $\phi\in \textbf{Force}(s)$.
\end{lemma}
\begin{proof}
The semantics of perdurant obligations is as follows : 
for $s \in [s_1 \dots d \ldots s_2]$, we have the condition:  
\textbf{(A)} \(\phi \in Force(s)\). Now suppose that, for a fixed formula $\Phi$ and fixed situations $S_1, \dots, D, \ldots D_2$, \( \textbf{OblgPerdurant}(\Phi,S_1, D, S_2)\) holds. From the right hand side of the abbreviation (\ref{OBLIGATION-Perdurance-1}), with appropriate substitutions, we get: 
\[(\forall s)[(S_1 \sqsubseteq D \sqsubseteq  s \sqsubseteq  S_2 )                        
	\supset \textbf{Oblg}(\Phi,s)].\]
Suppose that for a fixed $S$, we have 
$(S_1 \sqsubseteq D \sqsubseteq  S \sqsubseteq  S_2 $. Then, by the correlation (R1), $ S \in \textbf{Force}$. Henceforth, the  condition \textbf{(A)} is satisfied. 
$\Box$
\end{proof}

\begin{lemma}(\textbf{Violation of Perdurant Obligation})
If  the violation predicate $\textbf{violatedPerdurant}(\phi,s_1,d,s_2)$ holds, then \(\phi\) is a perdurant obligation in force in all situations $s$ in the  interval $[s_1 \dots d]$, that is, for all $s \in [s_1 \dots d]$, $\phi\in \textbf{Force}(s)$, and $\phi \not\in \textbf{State}(s)$.
\end{lemma}
\begin{proof}
For the violation of the perdurant obligation, from right hand side of the abbreviation~(\ref{Perdurance-obligation-violation-1}),  $\phi$ is not $\textit{true}$, meaning that $\phi  \in \textbf{Force}(s) $  but $\phi  \not\in \textbf{State}(s) $. 
$\Box$
\end{proof}


\section{Obligation Violation and Compensation}\label{Discussion-1}


\subsection{Compensation}

We first need to introduce the notion of compensation. Obligations that are violated are meant as penalty to be compensated by the violator. Intuitively, a compensation of a violation is a set of measures that the violator must take to make amend for the violation. Formally, a compensation $\textbf{Comp}$ is defined as a function ${\cal {L}}\mapsto 2^{\cal{L}}$, such that, for a given obligation $\phi$, $\textbf{Comp}(\phi)$ is the set of all obligations $\phi'$, called compensation obligations,  that come in force after $\phi$ has been violated. That is, $\textbf{Comp}(\phi) \in 2^{\cal{L}}$. 

An obligation $\phi$ is compensable if and only if $\textbf{Comp}(\phi)\not= \emptyset$, and for all $\phi' \in \textbf{Comp}(\phi)$, there is a $t \in \mathbb{N}$ such that  $\phi' \in \textbf{Force}(t)$; that is, a compensable obligation is associated with a non empty set of obligations that are in force at least any one time point. 

Finally, an obligation $\phi$ is compensated if and only if $\phi$ is violated and every $\phi' \in \textbf{Comp}(\phi)$ is either not violated or compensated. To keep matters simple, we assume that compensation obligations are never violated, thus avoiding (for now) the possibility of recursive compensation.   

\subsection{Compensation Action} 

An obligation violation must be countered by a compensation action that must be executed whenever it is possible, which is when an obligation violation has occurred; that is, when $violatedType()$ is true. This is done in a manner similar to natural actions of  Reiter~\cite{DBLP:conf/kr/Reiter96} that are forced to happen whenever their preconditions are met. We introduce a compensation action $execComp(t)$ which is similar to natural actions, and whose effect is to compensate violated obligations. 
Next, like for any action, we must provide an action precondition axiom for the newly introduced compensation action as follows:
\begin{align}\label{APA-EXECCOMP}
\begin{split} 
    Poss&(execComp(\phi,t))\equiv \\
    &(\exists s)[violatedAt(\phi,s)\land start(s)=t] \lor\\
    &(\exists s_1, s_2,t_1,t_2)[violatedPreemptive(\phi,s_1, s_2)    
        \land start(s_1) = t_1 \land \\
    &\hspace{2cm} start(s_2)= t_2 \land t \leq t_2]\lor\\
    &(\exists s_1, s_2)[violatedNonPreemptive(\phi,s_1, s_2)\land  
         start(s_1) = t_1 \land \\
    &\hspace{2cm}start(s_2)= t_2 \land (t_1 \leq t \leq t_2)]\lor\\
    &(\exists s_1, s_2)[violatedMaintenance(\phi,s_1, s_2) \land  
         start(s_1) = t_1 \land \\
    &\hspace{2cm} start(s_2)= t_2 \land t \leq t_2]\lor\\ 
    &(\exists s_1, d,s_2)[violatedPerdurant(\phi,s_1, d,s_2) \land 
         start(s_1) = t_1 \land \\
   &\hspace{2cm} start(s_2)= t_2\land (t_1 \leq t \leq d \leq t_2)]. 
\end{split}
\end{align}
Furthermore, we introduce a fluent $compensated(\phi_1,\phi_2,s)$ whose truth value is affected by the compensation action $execComp(t)$ and which intuitively means that the obligation $\phi_2$ compensates the obligation $\phi_1$ in situation $s$. 
Finally, we extend Reiter's notion of executability of situations to those situations that include the compensation action as follows:   
\begin{align}\label{EXECUTABLE-SITUATIONS}
\begin{split}
	executable(&s) \defeq \\
	  (\forall a,s')[do&(a,s') \sqsubseteq s \supset 
	     Poss(a,s') \land start(s') \leq time(a)] \land \\
	 (\forall a',s'',t)&[Poss(execComp(\phi,t),s'') \land 
	         do(a',s'') \sqsubseteq s \supset 
      	     time(a') < t]. 
\end{split}
\end{align}

The axiom \ref{EXECUTABLE-SITUATIONS} expresses in its first conjunct of the right-hand side the condition for the executability of non-natural actions, while the second conjunct adds an additional condition that characterizes the executability of the $execComp(\phi,t)$  action.

\section{Related Work}\label{Discussion}


Our obligation-producing actions in the Situation Calculus are similar to and a  substantial modification of those that were first introduced  in ~\cite{DBLP:conf/ecai/Demolombe04} and in~\cite{DBLP:conf/deon/DemolombeH04},  where the deontic accessibility relationship $O$ was introduced. 
In ~\cite{DBLP:conf/ecai/Demolombe04} and~\cite{DBLP:conf/deon/DemolombeH04}, authors ranked deontic alternatives in terms of their levels of ideality; they subsequently define the obligatory sentences as those that are true in all alternative situations with maximal ideality; and, finally, they give a successor state axiom for the fluent $O$. By contrast, our work simplifies the formalization by removing any use of situation idealities and by solely embedding the possible world semantics for SDL from~ \cite{Hintikka70} in the Situation Calculus. Furthermore, we expand Reiter's regression to reason about obligation-producing actions.

Another approach for incorporating deontic notions into the Situation Calculus is presented by Classen and Delgrande in~\cite{DBLP:conf/kr/ClassenD20}. 
In~\cite{DBLP:conf/kr/ClassenD20}, deontic assertions and modalities are expressed as constraints that subsequently  compiled into a Situation Calculus action theory which are used to reason about obligations. We differ from this approach by expressing obligations directly in the Situation Calculus so that there is no need of an extra compilation step.  

In~\cite{Parvizimosaed2020}, authors have proposed a contract ontology called {\it Symboleo} to systematically organize the main concepts of typical legal contracts. These concepts include: contract, asset, role, power, obligation, party, and event. The structure of the contract ontology that is made of these concepts is summarized by the authors of~\cite{Parvizimosaed2020} in a meta-model. Symboleo is a language for specifying smart contracts and reasoning about them using statecharts and Event Calculus. There is a  striking difference between the Symboleo framework and the framework that this paper presents: Symboleo models the obligation modality without conceptualising the kind of obligation types that this paper pursues.      

Event Calculus  may also be used for specifying obligations~\cite{DBLP:conf/ruleml/HashmiGW14}. The Situation Calculus, however, enjoys the key advantage of the existence of GOLOG ~\cite{DBLP:journals/jlp/LevesqueRLLS97}, a Situation Calculus-based programming language for defining complex actions in terms of a set of primitive actions axiomatized in the Situation Calculus. 


\section{Conclusion and Future Work}\label{Conclusion}

This paper used the Situation Calculus account of obligations in the Situation Calculus presented in ~\cite{Kalonji-Kiringa-Yeap} to provide a Situation Calculus formal account of the taxonomy of obligations given in~\cite{DBLP:conf/ruleml/HashmiGW14}. Our formalization of this taxonomy used non-Markovian theories of the Situation Calculus to model the various types of obligations.  We showed the correctness of the formalization and we mentioned related work pertaining to the formalisms used to give a formal account of obligations. 

This paper has introduced obligation-producing actions into the temporal Situation Calculus. The ensuing formalism allows us to enlarge the scope of dynamic domains that can be specified in the language of the Situation Calculus. 

One avenue for future work will be to construct logical theories called {\it basic contractual theories} to formalize legal contracts along the tradition set by\cite{reiter2001knowledge}. Those basic contractual theories provide the formal semantics of the corresponding legal contracts. 

Another avenue for future work will be to represent legal contracts as processes in the Situation Calculus; such processes lead to states where desirable properties holds that logically follow from the basic contractual theory representing those legal contracts. Our approach will provide one with an implementable specification, thus allowing one to automatically check many properties of the specification using an interpreter. The later will be GOLOG \cite{DBLP:journals/jlp/LevesqueRLLS97}, a Situation Calculus-based programming language for defining complex actions in terms of a set of primitive actions axiomatized in the Situation Calculus. We will use GOLOG to develop a framework for obligation-based programming which we will apply to smart contracts. 

 

\end{document}